# Magnetic field dependence of antiferromagnetic resonance in NiO

Zhe Wang,[1] S. Kovalev,[1] N. Awari,[1,6] Min Chen,[1] S. Germanskiy,[1]
B. Green,[1] J.-C. Deinert,[1] T. Kampfrath,[2,3] J. Milano[4,5] and M. Gensch[1]

[1] *Institute of Radiation Physics, Helmholtz-Zentrum Dresden-Rossendorf, 01328 Dresden, Germany*
[2] *Department of Physics, Freie Universitt Berlin, Arnimallee 14, 14195 Berlin, Germany*
[3] *Fritz Haber Institute of the Max Planck Society, Faradayweg 4-6, 14195 Berlin, Germany*
[4] *CONICET-CNEA, Centro Atómico Bariloche, (R8402AGP) San Carlos de Bariloche, Ríó Negro, Argentina*
[5] *Instituto Balseiro, Universidad Nacional de Cuyo, Centro Atomico Bariloche, (R8402AGP) San Carlos de Bariloche, Argentina*
[6] *Zernike Institute for Advanced Materials, University of Groningen, Nijenborgh 4, 9747 AG Groningen, The Netherlands*

(Dated: June 20th, 2018)

We report on measurements of magnetic field and temperature dependence of antiferromagnetic resonances in the prototypical antiferromagnet NiO. The frequencies of the magnetic resonances in the vicinity of 1 THz have been determined in the time-domain via time-resolved Faraday measurements after selective excitation by narrow-band superradiant terahertz (THz) pulses at temperatures down to 3 K and in magnetic fields up to 10 T. The measurements reveal two antiferromagnetic resonance modes, which can be distinguished by their characteristic magnetic field dependencies. The nature of the two modes is discussed by comparison to an eight-sublattice antiferromagnetic model, which includes superexchange between the next-nearest-neighbor Ni spins, magnetic dipolar interactions, cubic magneto-crystalline anisotropy, and Zeeman interaction with the external magnetic field. Our study indicates that a two-sublattice model is insufficient for the description of spin dynamics in NiO, while the magnetic-dipolar interactions and magneto-crystalline anisotropy play important roles.

*Introduction* – An antiferromagnet with two spin-sublattices and uniaxial anisotropy is a classical model for introducing the concept of antiferromagnetic resonance [1,2]. Within each sublattice the magnetic moments are aligned parallel, while between the sublattices they are coupled by an antiferromagnetic exchange interaction. A uniaxial magneto-crystalline anisotropy can give rise to an antiferromagnetic resonance of finite frequency even without external magnetic field [2]. The classical theory of antiferromagnetic resonance can be well extended to account for different types of magnetic anisotropy, sample shapes, and external fields [2,3,4]. Recently, antiferromagnetic materials have attracted considerable attention for applications in spintronic devices [5], owing to the possibility to switch antiferromagnetic order simply by current pulses with writing speeds extending from the kHz [6] to the THz range [7]. Antiferromagnets have further unique features [5]: they are quite prone to external perturbing magnetic fields and do not generate stray fields. Therefore, the exploration of antiferromagnets over a wide range of parameters such as temperature and external magnetic field is desirable.

Due to its simple crystallographic structure and well-known antiferromagnetic order forming well above room temperature, NiO is a representative example for understanding antiferromagnetism and the related applications. Above the Néel ordering temperature of $T_N$ = 523 K, NiO crystallizes in a NaCl-type centrosymmetric cubic structure with the point group $m3m$ (see Fig.1). Below $T_N$, the magnetic moments (spins) of the $Ni^{2+}$ ions are ferromagnetically ordered in the {111} planes. The exchange interactions between the spins lead to lattice contraction along one of the four equivalent <111> stacking directions [8,9], resulting in four equivalent crystallographic twin domains. Due to magnetic anisotropy, the spins are oriented along one of the three <11$\bar{2}$> axes (i.e. <11$\bar{2}$>, <$\bar{2}$11>, and <1$\bar{2}$1>, see Fig. 1) within the {111} planes, which corresponds to three equivalent spin domains in each crystallographic domain [10].

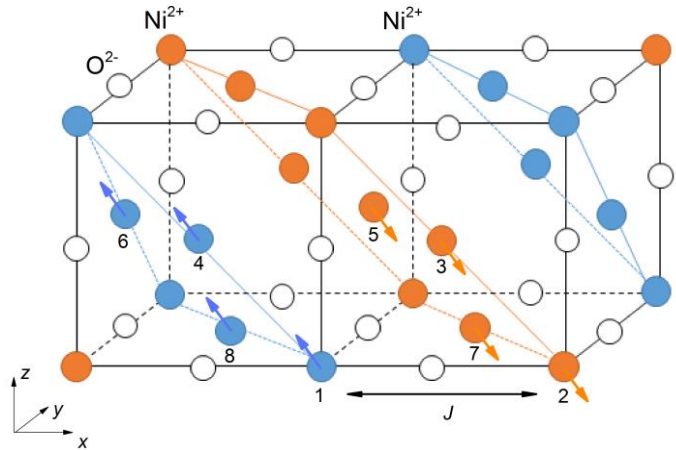

**Figure 1:** Illustration of the crystallographic and magnetic structure of NiO. Antiferromagnetic structure consists of ferromagnetically aligned spins in the (111) planes which are alternatingly stacked along the perpendicular direction. Below $T_N$, spins of the $Ni^{2+}$ ions are along the <**11$\bar{2}$**> directions, as indicated by the arrows. Eight simple cubic sublattices of Ni spins are indicated by the numbers marked at the origin of each sublattice. The dominant spin interaction is the superexchange between the next-nearest-neighbor $Ni^{2+}$ ions via the 180° $Ni^{2+}$–$O^{2-}$–$Ni^{2+}$ configuration as denoted by $J$.

The low-energy spin dynamics in NiO is characterized by antiferromagnetic spin-wave excitations as revealed by inelastic neutron scattering [11]. They were understood based on a two-sublattice antiferromagnetic model [2,11,12]. Further experimental studies, especially based on Raman and Brillouin spectroscopy, revealed five antiferromagnetic resonance modes close to the zero wave vector [13,14], thereby suggesting the magnetic structure to be more complex than a two-sublattice antiferromagnet. These observations were satisfactorily explained by an eight-sublattice model [13], which has been further extended to study the dependence of the three lowest-lying modes (below 0.5 THz) on an external magnetic field by Brillouin spectroscopy [15].

Experimentally, one high-energy mode at ~ 1THz has recently been shown to be coherently controllable by intense THz electromagnetic pulses at room temperature [16]. So far, however, it has remained unclear how the two highest-lying modes would evolve in an external magnetic field. Taking advantage of a narrow-band tunable superradiant THz source [17], we selectively excited these resonances and studied their temperature- and magnetic-field-dependent properties by recording transient Faraday rotation of a femtosecond near-infrared laser pulses. We show that while at room temperature only one antiferromagnetic spin resonance mode is present, which exhibits an evident nonlinear dependence on the external magnetic field, below 250 K one additional mode appears, characterized by its, in contrast, much weaker field dependence. We performed calculations based on an eight-sublattice model of the spin interactions in NiO, which can describe the observed field dependencies of the two high-frequency antiferromagnetic spin resonance modes.

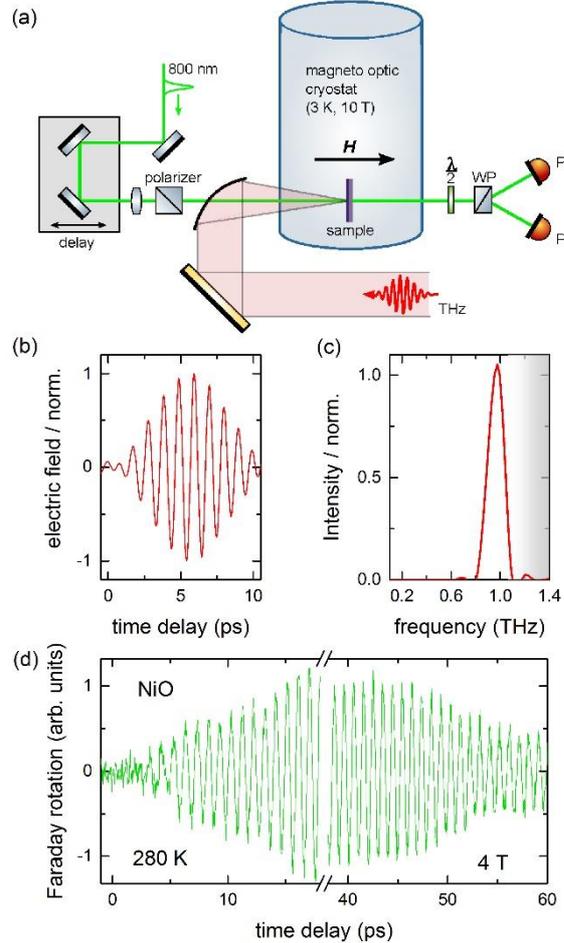

**Figure 2:** (a) Experimental endstation for THz pump Faraday probe experiments at the TELBE facility [17, 18]. Tunable narrow-band/multicycle THz pulses (red) are focused on the sample at normal incidence. Transient changes of the magnetization are probed by Faraday rotation of pulses from a collinear Ti:Sapphire based femtosecond laser (green) that can be timed to the THz pulses to better than 12 fs resolution [19]. $\frac{\lambda}{2}$, WP, and PD denote half-wave plate, Wollaston prism, and photodiode detector, respectively. (b) Electric field of the utilized 1 THz pump pulses and (c) the corresponding spectrum. (d) Typical transient Faraday measurement in an applied magnetic field of 4 T obtained at 280 K. The coherent oscillation of the magnetization is probed by the determination of the polarization change.

*Experimental details–* Figure 2(a) shows the experimental endstation for the THz pump Faraday rotation probe experiments at the TELBE facility based on a linear accelerator at the Helmholtz-Zentrum Dresden-Rossendorf [18]. Fourier-transform limited probe pulses were generated by Ti:Sapphire laser system with 100 fs pulse duration at 800 nm central wavelength. Timing jitter between laser system and the THz source was reduced to 12 fs based on a novel pulse-to-pulse data acquisition and analysis system [19]. Tunable narrow-band THz pulses with 20% bandwidth, 1 THz central frequency, and pulse energies of 1 μJ from a superradiant undulator were utilized to drive the antiferromagnetic spin resonance modes in crystalized NiO samples, which has been characterized previously as reported in Refs. [10,16]. The THz-pump pulses were measured at the sample position by electro-optical sampling in a 0.5 mm thick ZnTe crystal, see Figs. 2(c) and 2(d). The spectral density of the THz-pump pulses is depleted above 1.1 THz by water absorption lines. The transient magnetization change was recorded by measuring probe-pulse polarization-rotation angle at different time delays with respect to the pump pulse. A commercial magneto-optical cryostat (Oxford Instruments) was utilized for temperature dependent measurements down to 3 K and for applying an external magnetic field up to 10 T. The magnetic field was applied parallel to the incident laser beam and perpendicular to a (111) surface of a single-crystalline NiO sample. For this field orientation, the Ni moments within the different antiferromagnetic domains remain stabilized along the $<11\bar{2}>$ directions in finite fields.

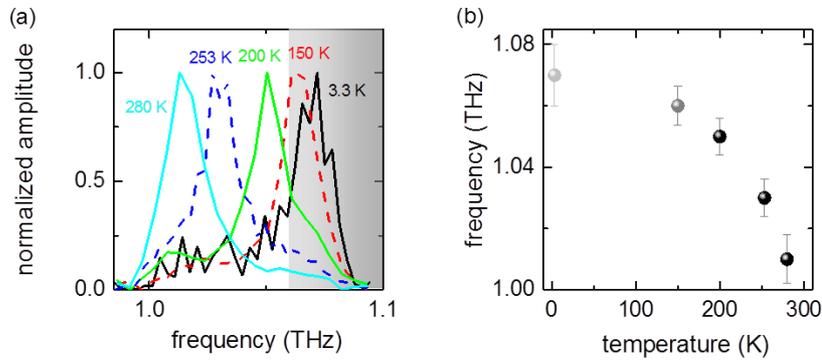

**Figure 3:** (a) Amplitude spectra obtained from Fourier transformation of the time-domain signals of Faraday rotation angle at various temperatures. The shaded area in the high-frequency limit indicates the spectral range where the THz radiation is strongly reduced by water absorption. (b) Temperature dependence of the peak frequencies in (a). Error bars indicate the apparent full widths at the half maxima.

*Results and discussion–* Measurements of the temperature-dependence at zero field have been performed for various temperatures between 3.3 and 280 K [see Fig. 3(a)]. At each temperature, the amplitude spectrum derived from Fourier transformation of the transient Faraday signal exhibits a single peak with well-defined position. The peak positions are shown in Fig. 3(b) as a function of temperature with the error-bars indicating the full widths at half maxima (FWHM) of the peaks. With decreasing temperature from 280 K, the peak position shifts to higher frequencies monotonically. This hardening of the peak frequency is consistent with previous measurements of the temperature dependence [1], which reflects the increase of the spontaneous magnetization of each sublattice with decreasing temperature [3]. It is worth noting that in our experiment, the higher-frequency components are affected by water absorption lines. Thus, for the low-temperature measurements the obtained resonance frequencies have larger uncertainty. For the same reason we could not resolve the higher-frequency mode that has been observed at 1.29 THz at low temperatures [20,21] (see Fig. 5, mode A).

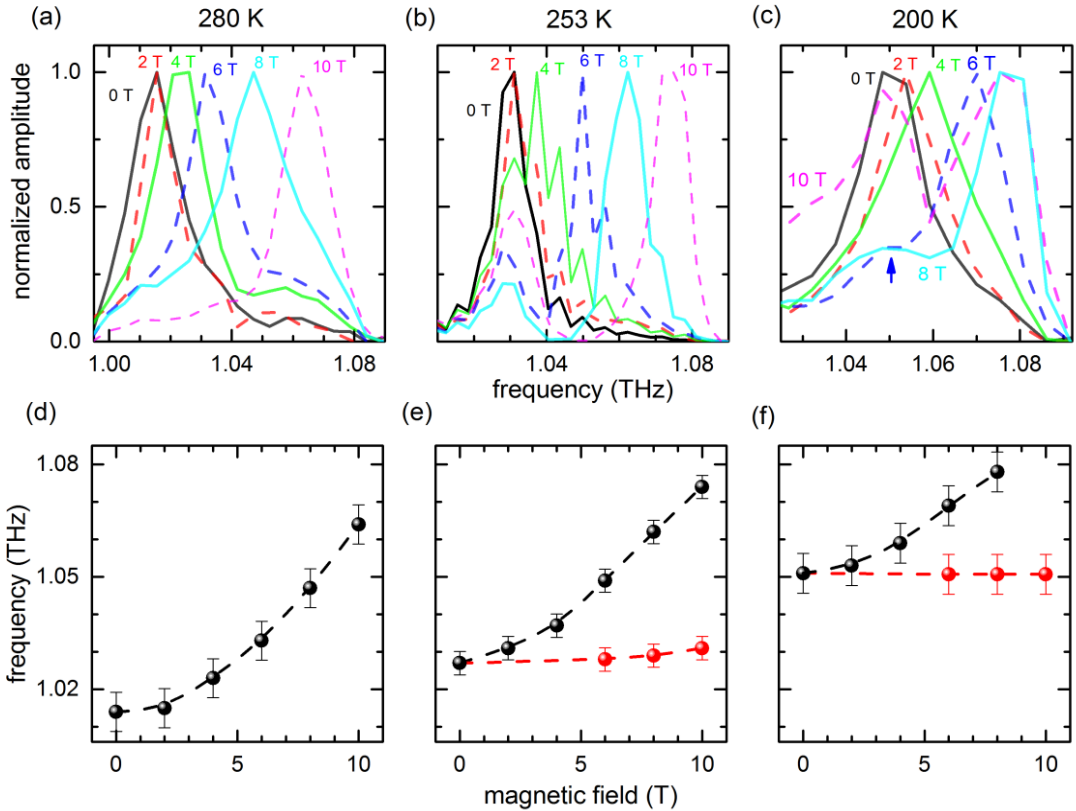

**Figure 4:** Magnetic field dependence at (a,d) 280 K, (b,e) 253 K and (c,f) 200 K. While only one mode is observed at 280 K, at the lower temperatures two modes are resolved. The qualitatively different field-dependencies of the two modes are in agreement with the predicted behaviors for mode A and mode B [see Fig. 5 and Eq.(1)]. In (d,e,f) the dashed lines are guides to the eyes.

While measurements of the temperature dependence have been performed in recent years [13,21], the magnetic-field dependence of the antiferromagnetic resonances at 1 THz has so far not been investigated. Figures 4 (a,b,c) show the experimentally obtained amplitude spectra for applied magnetic fields up to 10 T at 280 K, 253 K and 200 K, respectively. At 280 K, only one single peak is observed that continuously shifts to higher frequencies with increasing magnetic fields. In contrast, at 253 K and 200 K a second peak appears for higher magnetic fields. When plotting the peak positions over the magnetic field [Figs. 4(d,e,f)], it is clear that one resonance mode exhibits a significant nonlinear dependence on the external magnetic field while the other is almost field independent.

To understand the field-dependent behavior, we perform calculations of an eight-sublattice model following Ref. [15]. In contrast to the common two-sublattice model [11,12], which cannot explain our observations, we show that the eight-sublattice model correctly predicts not only the two antiferromagnetic modes around 1 THz, but also their characteristic field dependencies. In this model, the magnetic interactions comprises the antiferromagnetic exchange interactions $E_{exch}$, magnetic-dipole interactions $E_{dip}$, magnetic anisotropy $E_{ani}$, and a Zeeman interaction with an external magnetic field $E_{Zeeman}$, i.e.,

$$E = E_{exch} + E_{dip} + E_{ani} + E_{Zeeman} \tag{1}$$

with

$$E_{exch} = J(m_1 \cdot m_2 + m_3 \cdot m_4 + m_5 \cdot m_6 + m_7 \cdot m_8)$$
$$E_{dip} = D \sum_i \left( \sum_{j>i} m_i T_{ij} m_j \right),$$
$$E_{ani} = K \sum_i (m_{ix} m_{iy} m_{iz})^2,$$
$$E_{zeeman} = -g\mu_B H \cdot \sum_i m_i,$$

where $J > 0$ is the antiferromagnetic coupling constant, and $m_i$ is the unit magnetization vector at each sublattice site $j$ which is defined as $m_i = (\cos \phi_j \sin \theta_j, \sin \phi_j \sin \theta_j, \cos \theta_j)$. The dipolar interactions couple further the sublattices, which confines the spins in the {111} planes [15,22]. In the dipolar interaction term, the strength $D$ is determined by the magnetic moment of each Ni atom and the lattice constant, and the coupling tensors $T_{ij}$ have been explicitly given in Ref. [13]. For the anisotropy term, the magneto-crystalline constant $K < 0$ favors spins aligning along the <111> directions. A compromise with the stronger dipolar interactions leads to the orientation of spins close to the [11$\bar{2}$] direction [15].

The equilibrium values corresponding to the eight sublattices are found by direct minimization of the free-energy density given in Eq. (1). Following Refs. [23,24], a matrix is constructed from the second derivatives of the energy $E_{\theta(\phi)_i \phi(\theta)_j}$ with respect to magnetization angles $\theta_j$ and $\phi_j$, in which the matrix elements $B_{n,m}$ are given by [15]

$$B_{2i-1,2j-1} = E_{\theta_i \phi_j} / \sin \theta_j,$$
$$B_{2i,2j-1} = E_{\phi_i \phi_j} / \sin \theta_i \sin \theta_j,$$
$$B_{2i,2j} = -E_{\phi_i \theta_j} / \sin \theta_i,$$
$$B_{2i-1,2j} = -E_{\theta_i \theta_j}.$$

By solving eigenvalues of the matrix, frequencies of the antiferromagnetic modes are obtained as $\omega_k = i\gamma d_k / M$, where $\gamma$ is the gyromagnetic ratio (98 cm$^{-1}$/Oe for Ni) [25], $d_k$ stands for the eigenvalues, and $M$ is the saturation magnetization of each sublattice ($M = \frac{\mu_B}{a^3} = 128$ G).

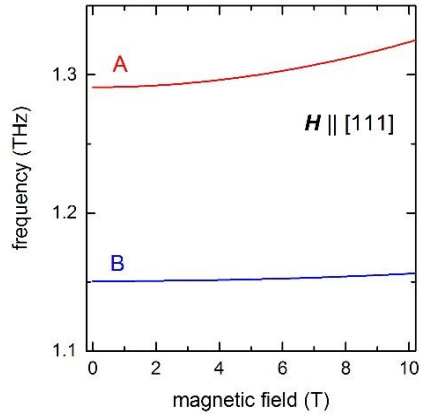

**Figure 5:** Field dependence of the higher-energy spin modes mode A at 1.29 THz (red) and mode B at 1.15 THz (blue) obtained from the eight-sublattice model [Eq. (1)], for the external magnetic field applied along the [111] direction.

The model of Eq. (1), essentially focusing on the zero-temperature spin dynamics, has properly described the experimentally-observed modes by Raman and Brillouin spectroscopy at the lowest temperatures [13], and the field dependencies of the three lower-lying modes [15]. According to this model, application of a high external magnetic field ($H > 2$ T) can lead to the instability of the spin domains in most situations. For example, if the external magnetic field is applied along the spin orientation of one spin domain, i.e. $H \parallel [11\bar{2}]$, the zero-field magnetic structure becomes unstable above ~ 1 T. A quite stable configuration is found for the external field applied along the [111] direction, which is exactly the orientation of a crystallographic domain that is perpendicular to the sample surface [10,16]. In this case, the spins are stabilized to be oriented along the <11$\bar{2}$> directions, meaning that the zero-field spin configuration remains stable at high fields. Thus, our theoretical results are intrinsic to a single spin domain, which are presented in Fig. 5 for the field dependence of the two highest-lying modes (mode A at 1.29 and mode B at 1.15 THz). We use the values of $J = 8.36 \times 10^8$ erg/cm$^3$, $D = -4.4 \times 10^4$ erg/cm$^3$, $K = 9 \times 10^4$ erg/cm$^3$ [15], and the value of Landé $g$-factor for the spin-1 Ni$^{2+}$ ions is taken as 2 [11]. The 1.15 THz mode is almost field-independent up to 10 T, because the oscillating magnetization components of this mode have larger components along the <1$\bar{1}$0> directions, perpendicular to the applied field. In contrast, the 1.29 THz mode evidently shifts to higher frequencies with increasing magnetic field. Comparing to the experimental observations, it is in an evident agreement on the field dependencies of the two resonance modes. Naturally, we can assume that the thermal effects [12] do not qualitatively alter the dependencies on an external magnetic field. Thus, we can assign the experimentally-observed mode with nonlinear field dependence (see Fig. 4) as the mode A of 1.29 THz obtained from the model calculations (see Fig. 5), while the other mode, observed at 253 and 200 K and almost field-independent up to 10 T, should correspond to the mode B of 1.15 THz. We note that, to fully describe the spin dynamics at elevated temperatures, a finite-temperature theory needs to be developed.

To conclude, we have studied the temperature and magnetic field dependence of the antiferromagnetic resonances in NiO. In high magnetic fields ($H > 6$ T) and at lower temperatures ($T \leq 253$ K), two different spin modes have been resolved with distinguished field dependencies. By performing calculations of an eight-sublattice model, the two modes are identified by their characteristic dependencies on the external magnetic fields. Thus, besides the antiferromagnetic exchange interactions of the Ni spins, our work has established that magnetic dipolar interactions and magneto-crystalline anisotropy are crucial for a proper description of the spin dynamics in the canonical antiferromagnet NiO. From the viewpoint of application, the existence of two spin modes with tunable frequency difference could provide more possibilities for the control over antiferromagnetic order through individual or combined resonant pumping of the two modes.

*Acknowledgements* – We would like to thank R. Bali, M. Fiebig, and M. Grimsditch for useful discussions. N.A. and M.G. acknowledge support from the European Commission's Horizon 2020 research and innovation programme, under Grant Agreement No. DLV 737038 (TRANSPIRE). B.G., S.K. and M.G. acknowledge support from the European Cluster of Advanced Laser Light Sources (EUCALL) project, which has received funding from the European Union Horizon 2020 research and innovation programme under Grant Agreement No. 654220. T.K. acknowledges the European Research Council for support through Grant No. 681917 (TERAMAG). J.M. acknowledges partial financial support from CONICET, ANPCyT and Universidad National de Cuyo.